\documentclass[10pt]{article}
\usepackage{latexsym,graphicx}
\usepackage{latexsym,graphicx}
\newcommand{\be}{\begin{equation}}
\newcommand{\ee}{\end{equation}}
\def\n{\noindent}
\catcode `\@=11 \catcode `\@=12
\begin{document}
\begin{center}
\large{\bf {A Plane-Symmetric Inhomogeneous Cosmological Model 
of Perfect Fluid Distribution with Electromagnetic Field I}} \\
\vspace{10mm}
\normalsize{Anirudh Pradhan\footnote{Corresponding author}, Prashant Kumar Singh$^2$, 
Anil Kumar Yadav$^3$} \\
\vspace{5mm} \normalsize{$^{1,2}$Department of Mathematics, Hindu
Post-graduate College,
 Zamania-232 331, Ghazipur, India} \\
\normalsize{$^1$E-mail: pradhan@iucaa.ernet.in}\\
\vspace{5mm} \normalsize{$^3$Department of Physics, K. N. Govt. Post-graduate 
College, Sant Ravidas Nagar (Gyanpur), Bhadohi-221 304, India} \\
\normalsize{abanilyadav@yahoo.co.in}\\
\end{center}
\vspace{10mm}
\begin{abstract}
A plane-symmetric inhomogeneous cosmological model of perfect fluid 
distribution with electro-magnetic field is obtained. The source of the magnetic 
field is due to an electric current produced along the z-axis. $F_{12}$ is the 
non-vanishing component of electromagnetic field tensor. To get a deterministic 
solution, we assume the free gravitational field is Petrov type-II non-degenerate. 
The behaviour of the electro-magnetic field tensor together with some physical 
aspects of the model are also discussed.
\end{abstract}
\smallskip
\n Key words : Cosmology, Electromagnetic field, Inhomogeneous solution .\\\
\n PACS: 98.80.Jk, 98.80.-k\\
\section{Introduction}
The standard Friedman-Robertson-Walker (FRW) cosmological model prescribes a homogeneous
and an isotropic distribution for its matter in the description of the present state of 
the universe. At the present state of evolution, the universe is spherically symmetric 
and the matter distribution in the universe is on the whole isotropic and homogeneous. 
But in early stages of evolution, it could have not had such a smoothed picture. Close 
to the big bang singularity, neither the assumption of spherical symmetry nor that of 
isotropy can be strictly valid. So we consider plane-symmetric, which is less 
restrictive than spherical symmetry and can provide an avenue to study inhomogeneities. 
Inhomogeneous cosmological models play an important role in understanding some essential 
features of the universe such as the formation of galaxies during the early stages of 
evolution and process of homogenization. The early attempts at the construction of such 
models have done by Tolman \cite{ref1} and Bondi \cite{ref2} who considered spherically 
symmetric models. Inhomogeneous plane-symmetric models were considered by 
Taub \cite{ref3,ref4} and later by Tomimura \cite{ref5}, Szekeres \cite{ref6}, Collins 
and Szafron \cite{ref7}, Szafron and Collins \cite{ref8}. Recently, Senovilla 
\cite{ref9} obtained a new class of exact solutions of Einstein's equation without 
big bang singularity, representing a cylindrically symmetric, inhomogeneous cosmological 
model filled with perfect fluid which is smooth and regular everywhere satisfying 
energy and causality conditions. Later, Ruis and Senovilla \cite{ref10} have separated 
out a fairly large class of singularity free models through a comprehensive study of 
general cylindrically symmetric metric with separable function of $r$ and $t$ as metric 
coefficients. Dadhich et al. \cite{ref11} have established a link between the FRW 
model and the singularity free family by deducing the latter through a natural and simple
in-homogenization and anisotropization of the former. Recently, Patel et al. 
\cite{ref12} presented a general class of inhomogeneous cosmological models filled with 
non-thermalized perfect fluid by assuming that the background space-time admits two 
space-like commuting killing vectors and has separable metric coefficients. Bali and 
Tyagi \cite{ref13} obtained a plane-symmetric inhomogeneous cosmological models of 
perfect fluid distribution with electro-magnetic field. Recently, Pradhan et al. 
\cite{ref14} have investigated plane-symmetric inhomogeneous cosmological models in 
different context.     
\par
The occurrence of magnetic fields on galactic scale is
well-established fact today, and their importance for a variety
of astrophysical phenomena is generally acknowledged as pointed
out by Zeldovich et al. \cite{ref15}. Also Harrison
\cite{ref16} has suggested that magnetic field could have a
cosmological origin. As a natural consequences, we should include
magnetic fields in the energy-momentum tensor of the early
universe. The choice of anisotropic cosmological models in
Einstein system of field equations leads to the cosmological
models more general than Robertson-Walker model \cite{ref17}. The
presence of primordial magnetic fields in the early stages of the
evolution of the universe has been discussed by several authors
\cite{ref18}$-$\cite{ref27}. Strong magnetic fields can be
created due to adiabatic compression in clusters of galaxies.
Large-scale magnetic fields give rise to anisotropies in the
universe. The  anisotropic pressure created by the magnetic
fields dominates the evolution of the shear anisotropy and it
decays slower than if the pressure was
isotropic\cite{ref28,ref29}. Such fields can be generated at the
end of an inflationary epoch \cite{ref30}$-$\cite{ref34}.
Anisotropic magnetic field models have significant contribution
in the evolution of galaxies and stellar objects. Bali and Ali
\cite{ref35} had obtained a magnetized cylindrically symmetric
universe with an electrically neutral perfect fluid as the source
of matter. Pradhan et al. \cite{ref36} have investigated
magnetized viscous fluid cosmological models in different context.
\par 
In this paper, we have obtained a new plane-symmetric inhomogeneous cosmological model 
of perfect fluid distribution with electromagnetic field. To get deterministic solution, 
we consider the free gravitational field is Petrov type-II non-degenerate in which the 
distribution is that of perfect fluid. The paper is organized as follows. The metric and 
the field equations are presented in Section 2. In Section 3, we deal with the solution 
of the field equations. Section 4 includes the physical and geometric features of the models. 
Finally the results are discussed in Section 5.
\section{The metric and field  equations}
We consider the metric in the form 
\begin{equation}
\label{eq1} ds^{2} = A^{2}(dx^{2} - dt^{2}) + B^{2} dy^{2} +
C^{2} dz^{2},
\end{equation}
where the metric potential $A$, $B$ and $C$ are functions of $x$ and $t$. 
The energy momentum tensor is taken as 
\begin{equation}
\label{eq2} T^{j}_{i} = (\rho + p)v_{i}v^{j} + p g^{j}_{i} + E^{j}_{i},
\end{equation}
where $E^{j}_{i}$ is the electro-magnetic field given by
Lichnerowicz \cite{ref37} as
\begin{equation}
\label{eq3} E^{j}_{i} = \bar{\mu}\left[h_{l}h^{l}(v_{i}v^{j} +
\frac{1}{2}g^{j}_{i}) - h_{i}h^{j}\right].
\end{equation}
Here $\rho$ and $p$ are the energy density and isotropic pressure respectively and 
$v^{i}$ is the flow vector satisfying the relation
\begin{equation}
\label{eq4} g_{ij} v^{i}v^{j} = - 1.
\end{equation}
$\bar{\mu}$ is the magnetic permeability and $h_{i}$ the magnetic flux
vector defined by
\begin{equation}
\label{eq5} h_{i} = \frac{1}{\bar{\mu}}~~ ^*F_{ji}v^{j},
\end{equation}
where $^*F_{ij}$ is the dual electro-magnetic field tensor defined
by Synge \cite{ref38} 
\begin{equation}
\label{eq6} ^*F_{ij} = \frac{\sqrt-g}{2}\epsilon_{ijkl} F^{kl}.
\end{equation}
$F_{ij}$ is the electro-magnetic field tensor and $\epsilon_{ijkl}$ is the Levi-Civita 
tensor density. The coordinates are considered to be comoving so that $v^{1}$ = $0$ 
= $v^{2}$ = $v^{3}$ and $v^{4}$ = $\frac{1}{A}$. We consider that the current is 
flowing along the z-axis so that $h_{3} \ne 0$, $h_{1} = 0 = h_{2} = h_{4}$. The only 
non-vanishing component of $F_{ij}$ is $F_{12}$. The Maxwell's equations
\begin{equation}
\label{eq7} 
F_{ij;k} + F_{jk;i} + F_{ki;j} = 0
\end{equation}
and
\begin{equation}
\label{eq8} 
\Biggl[\frac{1}{\bar{\mu}} F^{ij}\Biggr]_{;j} = J^{i}
\end{equation}
require that $F_{12}$ be function of $x$ alone. We assume that the magnetic 
permeability as a function of $x$ and $t$ both. Here the semicolon
represents a covariant differentiation. \\

The Einstein's field equations ( in gravitational units c = 1, G
= 1 ) read as
\begin{equation}
\label{eq9} R^{j}_{i} - \frac{1}{2} R g^{j}_{i} + \Lambda
g^{j}_{i} = - 8\pi T^{j}_{i},
\end{equation}
for the line element (1) has been set up as
\[
8\pi A^{2}\left( p + \frac{F^{2}_{12}}{2\bar{\mu}A^{2}B^{2}}\right) = - \frac{B_{44}}
{B} - \frac{C_{44}}{C} + \frac{A_{4}}{A}\left(\frac{B_{4}}{B} + \frac{C_{4}}{C}\right)
\]
\begin{equation}
\label{eq10} 
+ \frac{A_{1}}{A}\left(\frac{B_{1}}{B} + \frac{C_{1}}{C}\right) + \frac{B_{1}C_{1}}
{BC} - \frac{B_{4}C_{4}}{BC} - \Lambda A^{2},
\end{equation}
\begin{equation}
\label{eq11} 
8\pi A^{2}\left( p + \frac{F^{2}_{12}}{2\bar{\mu}A^{2}B^{2}}\right) = -\left(\frac{A_{4}}
{A}\right)_{4} + \left(\frac{A_{1}}{A}\right)_{1} - \frac{C_{44}}{C} + \frac{C_{11}}{C} 
- \Lambda A^{2},
\end{equation}
\begin{equation}
\label{eq12} 
8\pi A^{2}\left(p - \frac{F^{2}_{12}}{2\bar{\mu}A^{2}B^{2}}\right) = -\left(\frac{A_{4}}
{A}\right)_{4} + \left(\frac{A_{1}}{A}\right)_{1} - \frac{B_{44}}{B} + \frac{B_{11}}{B} 
- \Lambda A^{2},
\end{equation}
\[
8\pi A^{2}\left(\rho + \frac{F^{2}_{12}}{2\bar{\mu}A^{2}B^{2}}\right) = - \frac{B_{11}}
{B} - \frac{C_{11}}{C} + \frac{A_{1}}{A}\left(\frac{B_{1}}{B} + \frac{C_{1}}{C}\right)
\]
\begin{equation}
\label{eq13} 
+ \frac{A_{4}}{A}\left(\frac{B_{4}}{B} + \frac{C_{4}}{C}\right) - \frac{B_{1}C_{1}}
{BC} + \frac{B_{4}C_{4}}{BC} + \Lambda A^{2},
\end{equation}
\begin{equation}
\label{eq14} 
0 = \frac{B_{14}}{B} + \frac{C_{14}}{C} - \frac{A_{1}}{A}\left(\frac{B_{4}}{B} + 
\frac{C_{4}}{C}\right) - \frac{A_{4}}{A}\left(\frac{B_{1}}{B} + \frac{C_{1}}{C}\right), 
\end{equation}
where the sub indices $1$ and $4$ in A, B, C and elsewhere indicate ordinary 
differentiation with respect to $x$ and $t$, respectively. 
\section{Solution of the field equations}
Equations (\ref{eq10}) - (\ref{eq12}) lead to 
\[
\left(\frac{A_{4}}{A}\right)_{4} - \frac{B_{44}}{B} + \frac{A_{4}}{A}\left(\frac{B_{4}}{B} 
+ \frac{C_{4}}{C}\right) - \frac{B_{4}C_{4}}{BC} = 
\]
\begin{equation}
\label{eq15} 
\left(\frac{A_{1}}{A}\right)_{1} + \frac{C_{11}}{C} - \frac{A_{1}}{A}\left(\frac{B_{1}}{B} 
+ \frac{C_{1}}{C}\right) - \frac{B_{1}C_{1}}{BC} = \mbox{a (constant)}
\end{equation}
and 
\begin{equation}
\label{eq16} 
 \frac{8\pi F^{2}_{12}}{\bar{\mu}B^{2}} = \frac{B_{44}}{B} - \frac{B_{11}}{B} + 
\frac{C_{11}}{C} - \frac{C_{44}}{C}.
\end{equation}
Eqs. (\ref{eq10}) - (\ref{eq14}) represent a system of five equations in six unknowns 
$A$, $B$, $C$, $\rho$, $p$ and $\Lambda$. For the complete determination of these 
unknowns one more condition is needed. As in the case of general-relativistic 
cosmologies, the introduction of inhomogeneities into the cosmological equations 
produces a considerable increase in mathematical difficulty: non-linear partial 
differential equations must now be solved. In practice, this means that we must proceed 
either by means of approximations which render the non-linearities tractable, or we must 
introduce particular symmetries into the metric of the space-time in order to reduce the 
number of degrees of freedom which the inhomogeneities can exploit. In the present case, 
we assume that the metric is Petrov type-II non-degenerate. This requires that
\[
\left(\frac{B_{11} + B_{44} + 2B_{14}}{B}\right) - \left(\frac{C_{11} + C_{4} + 2C_{14}}
{C}\right) = 
\]
\begin{equation}
\label{eq17} 
\frac{2(A_{1} + A_{4})(B_{1} + B_{4})}{AB} - \frac{2(A_{1} + 
A_{4})(C_{1} + C_{4})}{AC}. 
\end{equation}
Let us consider that
\[
A = f(x)\lambda(t),
\]
\[
B = g(x)\mu(t),
\]
\begin{equation}
\label{eq18} 
C = g(x)\nu(t).
\end{equation}
Using (\ref{eq18}) in (\ref{eq14}) and (\ref{eq17}), we get
\begin{equation}
\label{eq19} 
\left[\frac{\frac{g_{4}}{g} - \frac{f_{1}}{f}}{\frac{g_{1}}{g}}\right] = 
\left[\frac{\frac{2\lambda_{4}}{\lambda}}{\frac{\mu_{4}}{\mu} + \frac{\nu_{4}}
{\nu}}\right] = \mbox{b (constant)}
\end{equation}
and
\begin{equation}
\label{eq20}
\frac{\frac{\mu_{44}}{\mu} - \frac{\nu_{44}}{\nu}}{\frac{\mu_{4}}{\mu} - 
\frac{\nu_{4}}{\nu}} - \frac{2\lambda_{4}}{\lambda} = 2\left(\frac{f_{1}}{f} - 
\frac{g_{1}}{g}\right) = \mbox{L (constant)}.
\end{equation}
Equation (\ref{eq19}) leads to
\begin{equation}
\label{eq21}
f = ng^{(1 - b)}
\end{equation}
and
\begin{equation}
\label{eq22}
\lambda = m(\mu \nu)^{\frac{b}{2}},
\end{equation}
where m and n are constants of integration. Equations (\ref{eq15}), (\ref{eq18}) 
and (\ref{eq20}) lead to 
\begin{equation}
\label{eq23}
\left(\frac{b}{2} - 1\right)\frac{\mu_{44}}{\mu} + (b - 1)\frac{\mu_{4}\nu_{4}}
{\mu \nu} = a
\end{equation}
and
\begin{equation}
\label{eq24}
(2 - b)\frac{g_{11}}{g} + (3b - 4)\frac{g^{2}_{1}}{g^{2}} = a.
\end{equation}
Let us assume 
\begin{equation}
\label{eq25}
\mu = e^{U + V}
\end{equation}
and 
\begin{equation}
\label{eq26}
\nu = e^{U - V}.
\end{equation}
Equations (\ref{eq20}), (\ref{eq25}) and (\ref{eq26}) lead to 
\begin{equation}
\label{eq27}
V_{4} = Me^{Lt + 2(b - 1)U},
\end{equation}
where M is constant. From equations (\ref{eq23}), (\ref{eq25}), (\ref{eq26}) and 
(\ref{eq27}), we have 
\begin{equation}
\label{eq28}
(b - 1)U_{44} + 2(b - 1)U^{2}_{4} - 2bM e^{Lt + 2(b -1)U}U_{4} - ML e^{Lt + 2(b -1)U}
= a.
\end{equation}
If we put $e^{2U} = \xi$ in equation (\ref{eq28}), we obtain
\begin{equation}
\label{eq29}
\frac{(b - 1)}{2}\frac{d^{2}\xi}{dt^{2}} - M \frac{d}{dt}(e^{Lt}\xi^{b}) = a\xi
\end{equation}
If we consider $\xi = e^{qt}$, then equation (\ref{eq29}) leads to
\begin{equation}
\label{eq30}
\frac{(b - 1)}{2}g^{2}e^{qt} - M\frac{d}{dt}(e^{Lt}e^{qbt}) = ae^{qt},
\end{equation}
which again reduces to
\begin{equation}
\label{eq31}
q = \frac{L}{1 - b}
\end{equation}
and
\begin{equation}
\label{eq32}
a = \frac{L(L + 2M)}{2(b - 1)}.
\end{equation}
Thus
\begin{equation}
\label{eq33}
U = \frac{Lt}{2(1 - b)}
\end{equation}
Equations (\ref{eq27}) and (\ref{eq33}) reduce to  
\begin{equation}
\label{eq34}
V = Mt + \log {N},
\end{equation}
where $N$ is an integrating constant. Eq. (\ref{eq24}) leads to 
\begin{equation}
\label{eq35}
g = \beta \cosh^{\frac{2 -b}{2(b - 1)}}{(\alpha x + \delta)},
\end{equation}
where
\begin{equation}
\label{eq36}
\alpha = \frac{\sqrt{2(3b - 4)(1 - b)}}{(2 - b)}
\end{equation}
and $\beta$, $\delta$ being constants of integration. Hence
\begin{equation}
\label{eq37}
f = n\beta \cosh^{\frac{b - 2}{2(b - 1)}}{(\alpha x + \delta)},
\end{equation}
\begin{equation}
\label{eq38}
\lambda = m e^{\frac{Ltb}{2(1 - b)}}, 
\end{equation}
\begin{equation}
\label{eq39}
\mu = e^{\frac{Ltb}{2(1 - b)} + Mt + \log{N}}, 
\end{equation}
\begin{equation}
\label{eq40}
\nu = e^{\frac{Ltb}{2(1 - b)} - Mt - \log{N}}.
\end{equation}
Therefore, we have
\begin{equation}
\label{eq41}
A = f \lambda =  m n \beta e^{\frac{Ltb}{2(1 - b)}}\cosh^{\frac{b -2}{2}}
{(\alpha x + \delta)},
\end{equation}
\begin{equation}
\label{eq42}
B = g \mu = N \beta e^{\left(\frac{L}{1 - b} + 2M \right)\frac{t}{2}} 
\cosh^{\frac{2 - b}{2(b - 1)}}{(\alpha x + \delta)},  
\end{equation}
\begin{equation}
\label{eq43}
C = g \nu = \frac{\beta}{N}e^{\left(\frac{L}{1 - b} + 2M \right)\frac{t}{2}} 
\cosh^{\frac{2 - b}{2(b - 1)}}{(\alpha x + \delta)}.
\end{equation}
By using the transformation
\[
X = x + \frac{\delta}{\alpha},
\]
\[
Y = y,
\]
\[
Z = z,
\]
\begin{equation}
\label{eq44}
T = t,
\end{equation}
the metric (\ref{eq1}) reduces to the form
\[
ds^{2} = K^{2} \cosh^{b - 2}{(\alpha X)} e^{\frac{LTb}{1 - b}}(dX^{2} - dT^{2}) + 
\]
\begin{equation}
\label{eq45}
G^{2} \cosh^{\frac{2 - b}{b - 1}}{(\alpha X)} e^{\left(\frac{L}{1 - b} 
+ 2M \right)T}dY^{2}   + H^{2}\cosh^{\frac{2 - b}{b - 1}}{(\alpha X)} 
e^{\left(\frac{L}{1 - b} - 2M\right)T}dZ^{2},
\end{equation}
where $K = mn\beta$, $G = N\beta$ and $H = \frac{\beta}{N}$.
\section{Some Physical and Geometric Features}
The physical parameters, pressure $(p)$ and density $(\rho)$, for the model 
(\ref{eq45}) are given by
$$
8\pi p = \frac{1}{K^{2}}e^{\frac{mnbT}{b - 1}}\cosh^{2 - b}{(\alpha X)}\Biggl[
\frac{(2 - b)^{2} \alpha^{2}}{4(b - 1)}\left\{1 + \frac{2 - b}{b - 1}
\tanh^{2}{(\alpha X)}\right\}
$$
\begin{equation}
\label{eq46}
 - \frac{L^{2}}{4(1 - b)^{2}} - M^{2}\Biggr] - \Lambda,
\end{equation}

$$
8\pi \rho = \frac{1}{K^{2}}e^{\frac{mnbT}{b - 1}}\cosh^{2 - b}{(\alpha X)}\Biggl[
\frac{(2 - b) \alpha^{2}}{2(b - 1)}\left\{\frac{b}{b - 1}\tanh^{2}{(\alpha X)}
- 1\right\}
$$
\begin{equation}
\label{eq47}
 + \frac{L^{2}(2b - 1)}{4(1 - b)^{2}} - M^{2} -\frac{ML}{(1 - b)}\Biggr] + \Lambda.
\end{equation}
The non-vanishing component $F_{12}$ of the electromagnetic field tensor is 
given by
\begin{equation}
\label{eq48}
F_{12} = \sqrt{\frac{\bar{\mu}}{8\pi}\frac{2ML}{(1 - b)}} G e^{\left(\frac{L}{1 - b} 
+ 2M \right)\frac{T}{2}}\cosh^{\frac{2 - b}{2(b - 1)}}{(\alpha X)},
\end{equation}
where $\bar{\mu}$ remains undetermined as function of $x$ and $t$ both. \\
The scalar of expansion $(\theta)$ calculated for the flow vector $(v^{i})$ is given by
\begin{equation}
\label{eq49}
\theta = \frac{L(b + 2)}{2K(1 - b)} e^{\frac{LbT}{2(b - 1)}}\cosh^{\frac{(2 - b)}{2}}
{(\alpha X)}
\end{equation}
The shear scalar $(\sigma^{2})$, acceleration vector $(\dot{v}_{i})$ and proper 
volume $(V^{3})$ for the model (\ref{eq45}) are given by
\begin{equation}
\label{eq50}
\sigma^{2} = \frac{(L^{2} + 12 M^{2})}{12 K^{2}}e^{\frac{LbT}{(b - 1)}}
\cosh^{(2 - b)}{(\alpha X)}, 
\end{equation}
\begin{equation}
\label{eq51}
\dot{v}_{i} = \left(\frac{1}{2}(b - 2)\alpha \tanh(\alpha X), 0, 0, 0\right),
\end{equation}
\begin{equation}
\label{eq52}
V^{3} = K^{2} G H e^{\frac{LT(b + 1)}{(1 - b)}}\cosh^{\frac{(b - 2)(2b + 1)}{(b - 1)}}
{(\alpha X)}.
\end{equation}
From equations(\ref{eq49}) and (\ref{eq50}), we have
\begin{equation}
\label{eq53}
\frac{\sigma^{2}}{\theta^{2}} = \frac{(L^{2} + 12 M^{2})(1 - b^{2})}{3L^{2}(b + 2)^{2}}
= \mbox{constant}.
\end{equation} 
The rotation $\omega$ is identically zero and the non-vanishing component of 
conformal curvature tensor are given by
\begin{equation}
\label{eq54}  
C_{(1212)} = \frac{1}{6K^{2}}e^{\frac{LbT}{(b - 1)}}\cosh^{(2 - b)}{(\alpha X)}
\left[3ML - 2M^{2} + b \alpha - \frac{L^{2}}{4b}\right], 
\end{equation} 
\begin{equation}
\label{eq55}  
C_{(1313)} = \frac{1}{6K^{2}}e^{\frac{LbT}{(b - 1)}}\cosh^{(2 - b)}{(\alpha X)}
\left[b \alpha - \frac{L^{2}}{4b} - 3ML - 2M^{2}\right], 
\end{equation} 
\begin{equation}
\label{eq56}  
C_{(2323)} = \frac{1}{3K^{2}}e^{\frac{LbT}{(b - 1)}}\cosh^{(2 - b)}{(\alpha X)}
\left[\frac{L^{2}}{4b} - b \alpha + 2M^{2} + \frac{ML}{(1 - b)}\right], 
\end{equation} 
\begin{equation}
\label{eq57}  
C_{(1224)} =\frac{ML}{2K^{2}}e^{\frac{LbT}{(b - 1)}}\cosh^{(2 - b)}
{(\alpha X)}.
\end{equation}
The dominant energy condition is given by Hawking and Ellis \cite{ref39}
$$
(i) ~ ~ ~ ~ \rho - p \geq 0
$$
$$
(ii) ~ ~ ~ ~ \rho + p \geq 0
$$
lead to
\[
e^{\frac{mnbT}{(b - 1)}}\Biggl[\frac{(2 - b)\alpha^{2}}{2(b - 1)}\left\{\frac{(b^{2} 
- 6b +4)}{2(1 - b)}\tanh^{2}{(\alpha X)} + \frac{b - 4}{2}\right\} 
\]
\begin{equation}
\label{eq58} 
+ \frac{L^{2}b}{2(1 - b)^{2}} - \frac{ML}{(1 - b)} \Biggr] + 2K^{2}\Lambda 
\cosh^{(b - 2)}{(\alpha X)} \geq 0
\end{equation}
and
\[
\frac{(2 - b)\alpha^{2}}{4(b - 1)^{2}}\left[(b^{2} - 2b + 4)\tanh^{2}{(\alpha X)} 
+ b(1 - b)\right] \geq 
\]
\begin{equation}
\label{eq59}
\frac{L^{2}}{2(1 - b)} + 2 M^{2} + \frac{ML}{(1 - b)}.
\end{equation}
\section{Conclusion}
We have obtained a new plane-symmetric inhomogeneous cosmological model of 
electro-magnetic perfect fluid as the source of matter. Generally the model 
represents expanding, shearing, non-rotating and Petrov type-II non-degenerate 
universe in which the flow vector is geodesic. We find that the model starts 
expanding at $T = 0$ and goes on expanding indefinitely. However, if $b < 0$ the 
process of contraction starts and at $T = \infty$ the expansion stops. For large 
values of $T$, the model is conformally flat and Petrov type-II non-degenerate otherwise. 
Since $\frac{\sigma}{\theta} = $ constant, then the model does not approach 
isotropy. The electromagnetic field tensor does not vanish when 
$L \ne 0$, $M \ne 0$, and $b \ne 1$. For large values of $T$ and $L + 2M(1 - b) < 0$ then 
$F_{12}$ tends to zero. 
\section*{Acknowledgements}
One of the authors (A. Pradhan) thanks Professor G. Date, IMSc., Chennai, India for 
providing facility where part of this work was carried out.

\end{document}